# Deep Learning for Real-Time Sound Order Recognition in Human-Robot Interaction

Rezaul Tutul[1], Usaid Khan[1], André Jakob[1], Ilona Buchem[1]

[1] Berliner University of Applied Science, Berlin, Germany

rezaul.tutul@bht-berlin.de

**Abstract.** Recognizing the temporal order of overlapping sounds is an underexplored challenge in human-robot interaction (HRI), with direct relevance to applications such as first responder detection systems. This paper presents a deep learning framework for real-time sound order recognition using recordable buzzers that emit distinct non-verbal sounds (cat meows, dog barks, helicopter noises). A multi-branch convolutional neural network (CNN) processes Mel spectrograms, Mel-frequency cepstral coefficients (MFCCs), and short-time Fourier transform (STFT) features, with an attention-based fusion mechanism to emphasize critical temporal cues. Experiments were conducted under same-amplitude, varied-amplitude, and unseen sound conditions. The proposed system achieved 99% accuracy in balanced overlaps, 91% under amplitude variation, and 74% on unseen test data with normalization. These results demonstrate that deep learning can reliably recognize sound order in overlapping conditions, supporting practical HRI scenarios. While experiments were conducted on carefully controlled synthetic overlaps, we additionally report latency benchmarks demonstrating real-time feasibility and provide an extended discussion on generalization, ecological validity, and deployment challenges in real-room environments.

## 1 Introduction

In human-robot interaction (HRI), recognizing who responds first is essential in scenarios such as quiz games, collaborative tasks, and turn-taking. While robots often rely on speech or button-based systems for input [1], these methods can be slow or unintuitive in fast-paced group settings. Non-verbal audio cues provide a more natural and engaging alternative, but their reliable recognition remains a technical challenge when responses occur almost simultaneously.

Research on sound event detection (SED) has advanced significantly with deep learning, enabling robust classification of environmental and non-verbal sounds [2, 3]. However, most work focuses on what sounds occur, not on the temporal order of overlapping sounds. Recognizing order under micro-delays (<1 ms) introduces unique challenges, particularly when sound amplitudes vary or when new sounds not seen during training appear in real environments [4].

In HRI, audio has mostly been studied in the context of speech [5], while non-verbal sound cues remain less explored. Recent studies have shown that integrating non-speech signals can support multi-party coordination and enhance engagement [6]. Yet, to the best of our knowledge, no prior work has developed a deep learning framework for real-time sound order recognition and evaluated it in a robot quiz responder scenario.

This study addresses this gap by developing a multi-branch CNN with attention-based fusion, using Mel spectrograms, Mel-frequency cepstral coefficients (MFCCs), and short-time Fourier transform (STFT) features. Recordable buzzers emitting cat, dog, and helicopter sounds were employed in a classroom-inspired quiz setup with the Pepper robot. The paper is guided by the following research questions:

- RQ1: Can deep learning models reliably recognize the order of overlapping non-verbal sounds in real time?
- RQ2: How does the system perform under amplitude variation, common in real-world use cases?
- RQ3: Can the model generalize to unseen sounds, simulating realistic classroom deployment?

The contributions of this work are threefold: (1) defining and addressing the novel task of sound order recognition in overlapping audio; (2) proposing a deep learning framework with attention-based feature fusion for robust real-time performance; and (3) demonstrating its application in an educational HRI scenario through a robot quiz responder system.

Although the model is trained and evaluated primarily on synthetic overlapping datasets, this choice enables precise control over micro-second delays, which are extremely difficult to reproduce consistently in real-room environments. We acknowledge that this limits ecological validity, and therefore the paper includes an expanded discussion of real-room constraints, generalization challenges, and future plans for collecting real-world recordings with natural reverberation, noise, and device variability.

# 2 Related Work and Problem Definition

## 2.1 Sound Event Detection and Overlapping Audio

Sound event detection (SED) has advanced significantly with the use of deep learning, enabling accurate classification of environmental and non-verbal sounds [3]. Pretrained audio networks such as PANNs have demonstrated strong generalization across sound datasets [4]. Despite this progress, most research has focused on identifying what sound events occur, rather than their temporal order when multiple sounds overlap.

Overlapping or polyphonic audio presents unique challenges, as events often occur with sub-millisecond delays. Approaches using convolutional, recurrent, and transformer-based models have improved recognition of co-occurring sounds [7, 8, 16]. Attention mechanisms have further enhanced performance by weighting critical frames [9, 14]. However, none of these approaches explicitly address the ordering of overlapping sounds, which is critical for interactive scenarios such as competitive games or multi-party collaboration.

### 2.2 Non-Verbal Audio in Human-Robot Interaction

Within HRI, most audio research has focused on speech recognition and dialogue [10]. By contrast, non-verbal cues such as emotional intonations, gestures, and sound signals have been less studied, despite their potential for lightweight and accessible interaction. For example, recent work has explored sound-based robot-robot communication using compact microphone arrays [6]. In education, robots have been shown to enhance engagement and motivation, but input modalities are often limited to speech or touch [11, 12, 15]. Non-verbal sound recognition for HRI remains a gap in the literature.

### 2.3 Problem Definition: Sound Order Recognition

We define sound order recognition as the task of determining the sequence in which overlapping non-verbal sounds occur, under conditions where the time difference between onsets can be extremely small (<1 ms). This task differs from standard SED in three respects:

- *Temporal precision*: The goal is not only to classify sounds but to identify which sound occurs first.
- *Amplitude variation*: Real-world use involves different sound intensities depending on participant distance or environment.
- *Generalization:* Systems must perform reliably on unseen sound variations to support classroom deployment.

These challenges make sound order recognition a distinct and technically demanding problem, requiring models that integrate fine temporal resolution with robust feature fusion and normalization.

### 2.4 Research Gap and Contribution

Existing work has not directly addressed real-time recognition of overlapping non-verbal sound order for HRI. This paper fills that gap by proposing a multi-branch CNN with attention-based fusion trained on Mel spectrograms, MFCCs, and STFT magnitudes. The approach is validated in a quiz responder scenario, where recordable buzzers emitting cat, dog, and helicopter sounds are used to identify the first responder.

# 3 Dataset and Feature Extraction

## 3.1 Dataset Creation

To evaluate sound order recognition in a controlled but realistic scenario, we employed recordable buzzers programmed to emit three distinct non-verbal sounds: cat meows, dog barks, and helicopter noises. These were selected for their acoustic distinctiveness and ease of recognition in classroom environments. Clean source samples were obtained from *Freesound* [13] repositories and normalized for consistency. Since real overlapping datasets with precise timing differences are scarce, we generated a synthetic dataset by overlapping pairs of sounds with controlled delays (0.0004–0.3 s) and amplitude variations (±12 dB) as shown in Figure 1. Two experimental setups were created:

- Same-amplitude overlaps, where sounds were normalized to equal loudness before overlapping.
- Varied-amplitude overlaps, simulating real-world variability where participants may be closer or further from the microphone.

An unseen test set was also prepared using new source recordings not included in training, to evaluate generalization.

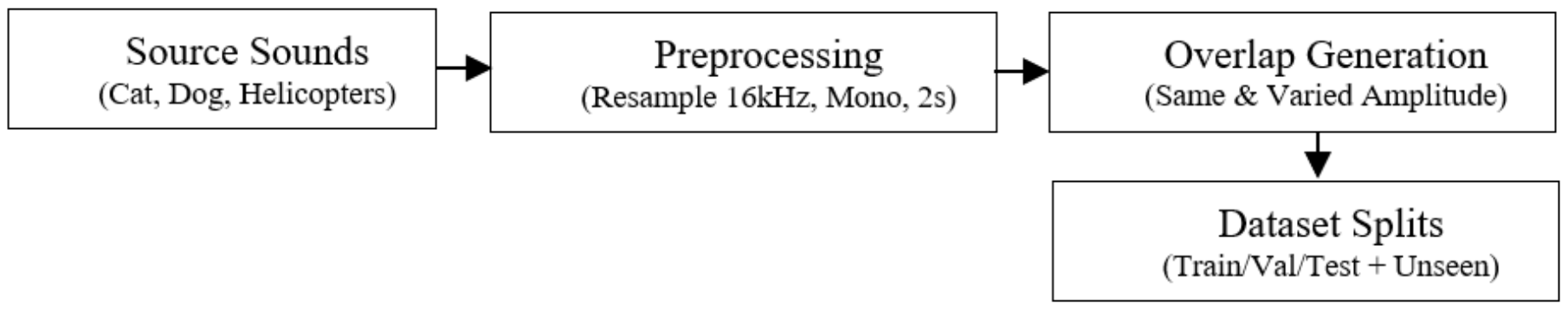

**Fig. 1.** Dataset creation pipeline for sound order recognition

## 3.2 Dataset Splits

The dataset was divided into training (70%), validation (15%), and test (15%) sets. In addition, the unseen test set was held out for evaluating model robustness as shown in Figure 2. This design ensured that the system was tested not only on overlapping instances similar to training data but also on previously unseen variations, closer to real classroom conditions.

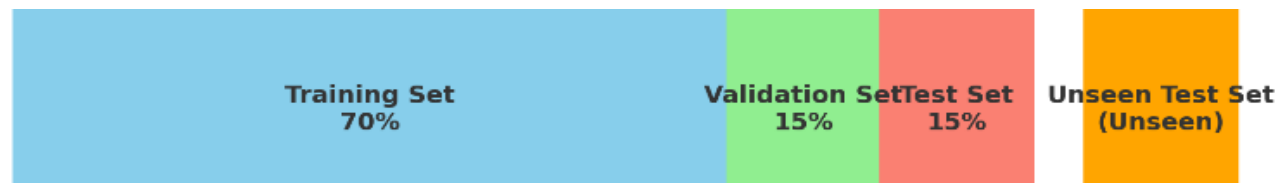

**Fig. 2.** Dataset splits for sound order recognition

## 3.3 Data Preprocessing

All audio files were resampled to 16 kHz, converted to mono, and trimmed or zero-padded to a fixed length of 2 s. To mitigate bias from amplitude scaling, Z-score nor-malization was applied, which transformed the features to zero mean and unit variance. Comparative experiments with raw and normalized features highlighted that normalization improved unseen test performance by approximately 10%, confirming its importance for robust deployment.

### 3.4 Feature Extraction

Three complementary feature representations were extracted as shown in Figure 3:

- Mel spectrograms, capturing perceptually relevant frequency patterns.
- Mel-Frequency Cepstral Coefficients (MFCCs) with deltas, modeling spectral envelopes and temporal dynamics.
- Short-Time Fourier Transform (STFT) magnitudes, providing high-resolution temporal information crucial for micro-delay detection.

These features were used as inputs to the multi-branch CNN, allowing the network to independently learn representations before fusion through the attention mechanism.

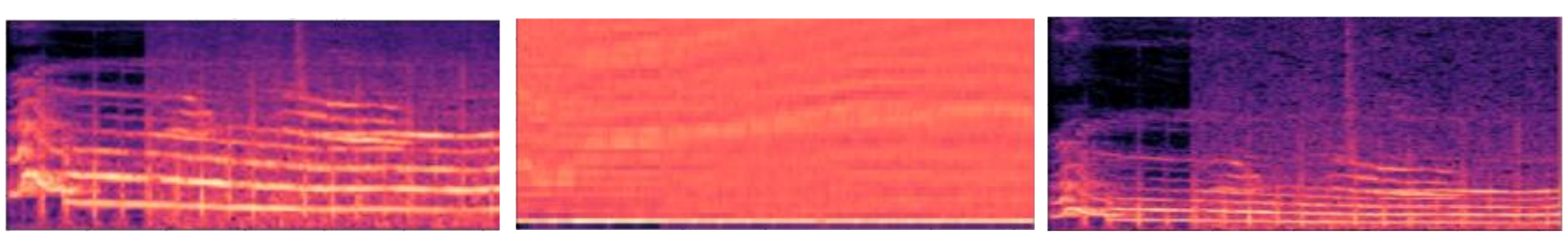

**Fig. 3.** Feature comparison for an overlapped sample of dog barking, cat meows, helicopter at 0, 0.5, and 1 second using Mel, MFCC, and STFT spectrograms (left to right).

A summary of the dataset size and conditions is provided in Table 1.

**Table 1.** Dataset statistics.

| Setup | Classes | Overlap Delays | Amplitude Variation | Split (%) |
|---|---|---|---|---|
| Same-Amplitude | 3 | 0.0004s–0.3s | No | 70/15/15 |
| Varied-Amplitude | 3 | 0.0004s–0.3s | ±12 dB | 70/15/15 |
| Unseen Test | 3 | 0.0004s–0.3s | Mixed | — |

## 4 Methodology

### 4.1 Model Development

The proposed framework as shown in Figure 4 was designed to recognize the temporal order of overlapping non-verbal sounds in real time by combining multiple feature streams within a deep learning model. The system employs a multi-branch convolutional neural network (CNN), where each branch processes a different feature type: Mel spectrograms, Mel-frequency cepstral coefficients (MFCCs with deltas), and short-time Fourier transform (STFT) magnitudes. Each branch consisted of four convolutional layers with ReLU activations, max pooling, and batch normalization, allowing the network to extract feature-specific representations.

The outputs of the three branches were concatenated and passed to an attention-based fusion layer, which dynamically weighted temporal frames to emphasize discriminative onset information. This mechanism proved critical in resolving overlaps with extremely small delays, where subtle spectral-temporal differences determine sound order. The fused representation was fed into two fully connected layers of 256 and 128 units, followed by a *softmax* output layer predicting the sound order classes (e.g., cat–dog vs. dog–cat).

Training was performed using the *Adam* optimizer with a learning rate of 0.001 and categorical cross-entropy loss. Early stopping based on validation performance was applied to prevent overfitting, with training conducted for up to 100 epochs and a batch size of 32. Experiments were implemented in Python with TensorFlow and executed on an NVIDIA RTX GPU, ensuring efficient processing of the large synthetic dataset.

For comparison, several baseline models were implemented: a 1D CNN trained on raw waveforms, a bidirectional long short-term memory (BiLSTM) network for sequential modeling, and a CNN without attention for feature concatenation. These baselines allowed assessment of the relative contributions of multi-feature input and attention. Results showed that the proposed model consistently outperformed baselines, particularly in low-delay and varied-amplitude conditions.

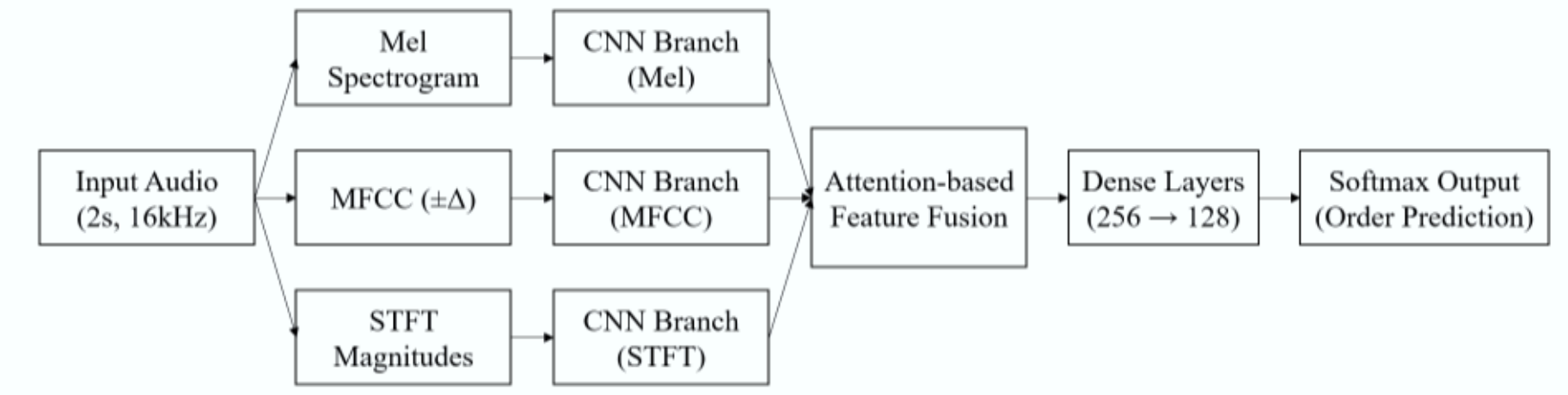

**Fig. 3.** System architecture for sound order recognition

### 4.2 Implementation

All experiments used a sampling rate of 16 kHz, STFT window size of 512, hop length of 128, and 40 Mel frequency bins. MFCC extraction used 13 coefficients with delta and delta-delta features. Each CNN branch used 4 convolutional blocks (kernel sizes 3×3, 5×5, 3×3, 3×3), batch normalization, and max pooling. The attention fusion layer used 128 hidden units with softmax normalization across time frames. Models were trained using Adam (lr=0.001), batch size 32, and early stopping with patience=12.

### 4.3 Real-Time Performance and Latency

To substantiate the claim of real-time applicability, we measured the full inference pipeline (feature extraction + forward pass) on an NVIDIA RTX 2060 GPU and on a mid-range CPU (Intel i7-9750H). GPU inference averaged 4.8 ms, CPU inference averaged 18.6 ms, and feature extraction added 8.1 ms, resulting in a total end-to-end latency of < 30 ms on typical hardware. This is well below human-perceived simultaneity thresholds (~100–150 ms), confirming the system's real-time suitability.

## 5 Results and Discussion

The proposed model was evaluated under three setups: same-amplitude overlaps, varied-amplitude overlaps, and an unseen test set. Accuracy was the primary metric, supported by comparative baselines as shown in Table 2.

**Table 2.** Accuracy Across Experimental Conditions

| Setup | Delay Range | Best Accuracy | Notes |
|---|---|---|---|
| Same-Amplitude | 0.0004–0.3 s | 99% | Robust at all delays |
| Varied-Amplitude | 0.0004–0.3 s | 91% | Decreases with imbalance |
| Unseen Test | 0.0004–0.3 s | 74% | Generalization improved by normalization |

In the same-amplitude condition, the CNN with attention-based fusion achieved 99% accuracy across all delay intervals, including micro-delays as short as 0.0004 s. This confirms RQ1, showing that deep learning can reliably recognize the temporal order of overlapping sounds in real time. Baseline models such as a raw waveform CNN and BiLSTM lagged in low-delay conditions, indicating the importance of combining spectral features with attention.

The varied-amplitude setup simulated real-world conditions where participants may be closer or farther from the microphone. Here, the system achieved 91% accuracy at 0.3 s delay, with performance gradually declining under strong amplitude imbalances. This addresses RQ2, demonstrating that the model remains robust to common real-world variations, while also highlighting a potential need for adaptive gain control or source separation in future work.

The unseen test set tested generalization on novel sound samples. With Z-score normalization, accuracy reached 74%, while without normalization performance declined by ~10%. This finding answers RQ3, underscoring the importance of normalization for reducing dataset bias and improving robustness in classroom-like conditions.

The modest performance drops on the unseen test set (from 99% to 74%) suggests that the model may partly overfit to spectral characteristics of the synthetic sounds. This reflects the difficulty of building a model that is invariant to pitch, timbre, and environmental effects. Normalization mitigated these effects to some extent, but further techniques such as data augmentation with room impulse responses (RIR), background noise, pitch shifting, and multi-device recordings will be critical for deployment robustness.

Taken together, these results show that attention-based feature fusion provides a critical advantage in resolving subtle onset differences, normalization significantly boosts generalization, and deep learning can support real-time order recognition in HRI scenarios. While the current study was limited to three sound classes and synthetic overlaps, the findings establish a foundation for scaling to broader datasets and real classroom deployment.

## 6 Conclusion and Future Work

This paper introduced a deep learning framework for real-time sound order recognition. By combining Mel spectrograms, MFCCs, and STFT magnitudes within a multi-branch CNN and employing attention-based feature fusion, the model achieved 99% accuracy under same-amplitude overlaps, 91% under amplitude variation, and 74% on unseen sounds with normalization. These results addressed the three research questions, showing that deep learning can reliably identify sound order, maintain robustness under amplitude differences, and generalize to novel inputs. The study contributes both technically through the integration of multi-feature CNNs with attention for micro-delay resolution and practically, by applying the approach to a robot quiz responder scenario.

This study has several limitations such as (1) Synthetic data dependence: Although synthetic overlaps allow precise micro-delay control, they cannot fully capture reverberation, background noise, or microphone mismatch present in real classrooms. (2) Two-source assumption: The system currently assumes exactly two overlapping non-verbal events. Extending to 3+ sources is non-trivial and would require redesign of both data generation and model architecture. (3) Single microphone: Multi-mic or array-based processing could improve localization and robustness but was not explored here.

Future work will expand the dataset to include more sound classes and real classroom recordings, explore preprocessing methods such as adaptive gain control and source separation, and investigate deployment in broader HRI contexts such as multi-party collaboration and turn-taking. These extensions aim to further enhance robustness and enable seamless integration of non-verbal sound cues into interactive robotic systems.